\documentclass[3p,times,twocolumn]{elsarticle}
     
\journal{Internet of Things}
       
\usepackage[export]{adjustbox}
\usepackage{textcomp}
\usepackage{times}
\usepackage{enumerate}
\usepackage{amsmath}
\usepackage{multirow}
\usepackage[T1]{fontenc}
\usepackage{color}
\usepackage{float}
\usepackage{array} 
\usepackage{multirow}
\usepackage[hyphens]{url}

\usepackage{blindtext} 
\usepackage{enumitem}
\usepackage{color}
\usepackage{afterpage}
\usepackage{subfig}
\usepackage{graphicx}
\usepackage[font=normal]{caption}

\usepackage{geometry}
\usepackage{siunitx}
\usepackage{amssymb,amsmath}
\usepackage{algorithm}
\usepackage{algpseudocode}
\usepackage{balance}
\usepackage{listings}
\usepackage{breqn}
\usepackage[utf8]{inputenc}

\newcolumntype{L}[1]{>{\raggedright\let\newline\\\arraybackslash\hspace{0pt}}m{#1}}
\newcolumntype{C}[1]{>{\centering\let\newline\\\arraybackslash\hspace{0pt}}m{#1}}
\newcolumntype{R}[1]{>{\raggedleft\let\newline\\\arraybackslash\hspace{0pt}}m{#1}}
\begin{document}

\begin{frontmatter}

\title{A Power-Efficient Audio Acquisition System for Smart City Applications}

\author[mymainaddress]{Evan Fallis}
\author[mymainaddress]{Petros Spachos\corref{mycorrespondingauthor}}
\cortext[mycorrespondingauthor]{Corresponding author}
\ead{petros@uoguelph.ca}
\author[mymainaddress]{Stefano Gregori}

\address[mymainaddress]{School of Engineering, University of Guelph, Guelph, Ontario, Canada}

\begin{abstract}
    Acoustic noise has adverse effects on human activities. Aside from hearing impairment and stress-related illnesses, it can also interfere with spoken communication, reduce human performance and affect the quality of life. As urbanization is intensifying, the potential benefits of reducing noise pollution in smart-city environments are extensive.  Noise levels can be collected and analyzed using a wireless sensor network which can monitor the noise level by using microphones. However, every wireless system struggles in terms of the battery requirements needed for continuous data collection and monitoring. In this paper, the design of a testbed for a smart microphone system is presented. To save power, a microcontroller and an Analog-to-Digital Converter (ADC) dynamically switch between high and low power modes in response to environmental noise. Specifically, the high powered components are triggered by a spike in the acoustic noise level. Three wireless technologies, WiFi (2.4 GHz), Bluetooth Low Energy (BLE) 4.0 and Zigbee were examined. According to the results, the power consumption of a node can be lowered by 97\% when idle based on the testbed.
\end{abstract}


\end{frontmatter}

\section{Introduction}
    A common problem in urban cities is acoustic noise pollution~\cite{khan}. Noise pollution is an abundance of noise in a concentrated area that can harm humans both physically and mentally~\cite{shepard}. To alleviate this problem, the first step is to understand and analyze what acoustic noise citizens are being exposed to daily. Smart cities enable the introduction of new technology to the public through different services that optimize resource usage, transportation, utilities and more. In a smart city, large amounts of data are collected and utilized~\cite{xu}. This is done to predict, analyze, and observe trends that happen throughout the city based on various sensors.

    Audio data is an example of sensory information being collected. The amount of human comfort in different locations within a city can be found by analyzing noise levels collected by multiple microphones. By creating a tool that can visualize the levels of acoustic noise around the city, its people are given the ability to react and adjust accordingly. In future iterations, it may also be possible to show the type of sounds present if a neural network is utilized~\cite{shu}. An example is the classification of different sounds such as human voice, traffic, construction, or air transport that is currently present in any school or hospital. Mapping the amount of sound classified as human voice within an area grants the ability of the public to observe the amount of human presence at the time. This can also help citizens to determine if an area within the city such as a park or restaurant is comfortable to visit at a specific time in the day.
    
    Analyzing the audio data in an urban environment can become complicated depending on the location within the city~\cite{rong}. There is a variety of challenges such as privacy and placement~\cite{christin}. Privacy has always been a big issue with regard to the collection of data in a public setting. An invasion of privacy could be claimed by anyone who feels uncomfortable about the idea of the voices being recorded.
    
    Another challenge is the reliability of the data, which depends on its accuracy in modeling the noise within an area. Data is considered reliable if the extraneous variables that could skew the noise model of the area are minimal. A major concern is the power consumption of various microphones and analog-to-digital converters (ADCs) when collecting audio data~\cite{shehzad}. This is especially true when the microphone is set to sample throughout the day, meaning that it will have minimum downtime. Many systems try to solve this problem by sampling at different points in the day rather than being continuous. Although this is a good method for saving energy, it can miss key information due to the scheduled sampling.
    
    The work presented relates to a framework for reducing power consumption in Micro-Electro-Mechanical System (MEMS) microphone ADCs for an acoustic environmental monitoring system. A testbed to prove the power savings has been developed. The contributions of this paper are as follows:
    
    \begin{itemize}
        \setlength\itemsep{0.5em}
        \item{A testbed for environmental sound detection using a microphone was designed. The microphone was selected after experimentation on the energy requirements and the accuracy of four different microphones.}
        \item{An amplifier and a threshold circuit were designed for use with audio collection systems.}
        \item{Three wireless technologies were used to examine the effectiveness of the proposed energy-saving scheme. Specifically, WiFi (2.4 GHz), Bluetooth Low Energy (BLE) 4.0 and Zigbee were used.}
    \end{itemize}
    
    The results are compared to a reference system to prove the viability and demonstrate the improvements. The rest of the paper is organized as follows; Section~\ref{related} reviews the related works and Section~\ref{experiments} discusses the proposed system architecture. Section~\ref{procedure} details the procedure involved and the results found during analysis are presented in Section~\ref{results}. The conclusion is in Section~\ref{conclusion}.

\section{Related Work}\label{related}
    In recent years, many research works involve wireless sensor networks for acoustic noise collection~\cite{peckens, risojevic, mydlarz, alsina, gubbi, noriega, hakala, kivela, santini, filipponi}. Their main focus is the development of systems that have low enough power to reliably operate with little to no human intervention. 

    A promising implementation has been developed in~\cite{peckens}. The work focuses on developing an ultra low power sensor node that is designed to wake up periodically to collect data, it was programmed to wake for ten minutes every hour. To accomplish this, an ultra-low-power microcontroller was used along with an XBee. The total power when transmitting was 810 mW while the transmitting power was 132 mW. A handheld Sound Pressure Level (SPL) meter was used to verify the accuracy of the audio level collection. Although this prototype may be sufficient to get the overall average noise level during a day, it can easily miss many important events since it is only awake less than 20\% of the time.
    
    The work in~\cite{risojevic} presents a node that collects audio level information as well as other types of data. It is designed to wake up based on a timer interrupt or when the accelerometer is triggered. The method of wireless transmission involves the Zigbee protocol to ensure low power. The model was verified by measuring the audio level of varying sinusoidal waves, ranging from 31.5 Hz to 16 kHz. This was recorded with an MI 6201 MULTINORM sound level meter. The average power consumption of the prototype is 45 mW. This is extremely low when compared to most devices, allowing operation for up to 7 days. This prototype is a good starting point but requires a longer operating time to reduce the need for continuous maintenance.
 
    \begin{figure*}[t!]
        \centering
        \includegraphics[width=0.85\textwidth]{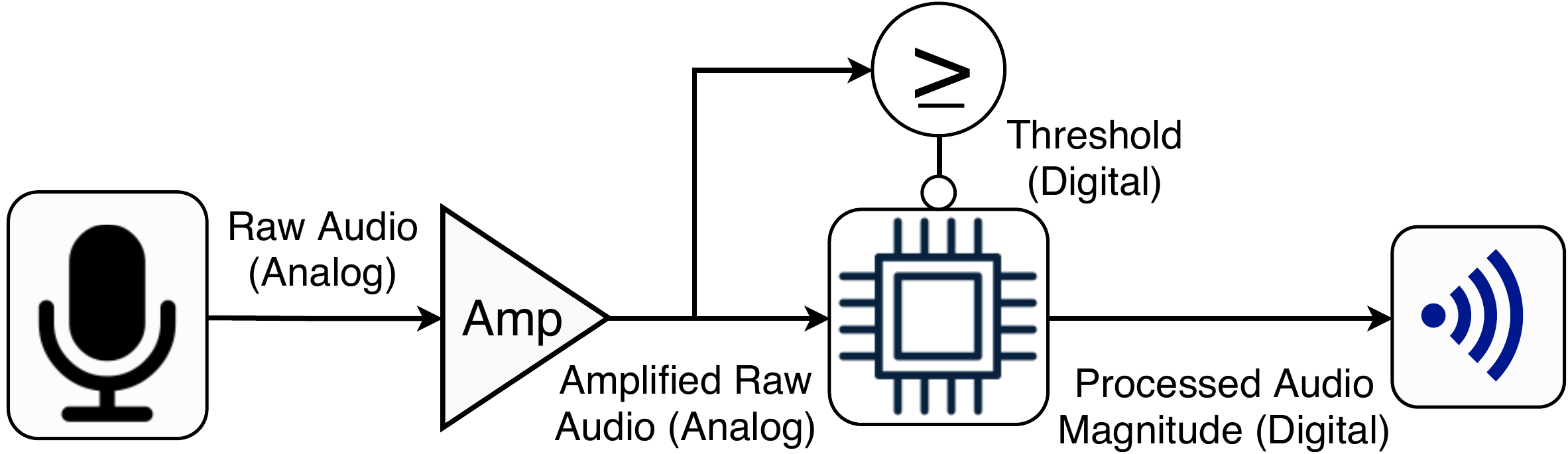}
        \caption{System overview.}
        \label{testbed}
    \end{figure*}
    
    The work in \cite{mydlarz} features a prototype using a mini PC. Although the system is not fully wireless, it is a good solution for indoor environments. It was verified using a calibrated sound level meter using an A-weighted filter. In \cite{alsina}, a prototype was built for use with city buses. This had a power supply unit connected to the city bus and used 3G as well as GPS to wirelessly transmit data along with a location stamp. As described in the article, one challenged faced was attempting to filter out the noise generated from the bus, this presents a big problem with the accuracy of the data if it is meant to represent noise level from a human perspective. 
    
    A pilot study was presented in~\cite{gubbi}, which involved an embedded system to collect acoustic noise. The power consumption was not specified but the accuracy of collected data was discussed extensively. A standard sound level meter was used to calibrate the system and verify the analog filters. A noise monitoring node based around a Raspberry Pi 2B and USB microphone was developed in~\cite{noriega}. A sound level meter was used to verify all measurements and power over Ethernet supplied each node with energy. Although not wireless, it was still considered a flexible solution for an acoustic sensor network.
    
    In~\cite{hakala}, a low-cost sensor network was developed to target the problem of noise pollution. It has an operating power of 89 mW when collecting data. A sound level meter was used to calibrate the system, ensuring accurate measurements. Despite the low power, the prototype can run for less than 4 days, which is not very realistic for a full implementation. A similar implementation was followed in~\cite{kivela}. The authors in~\cite{santini} show preliminary results of a wireless noise collection system, mainly discussing the accuracy of this method and giving recommendations for future designs. In~\cite{filipponi}, the authors focus on a prototype involving low power listening. They mainly focus on the networking protocols involved and recommend the collection tree protocol.

    Smartphones have been considered for noise collection via crowdsourcing. The authors in~\cite{zappatore} developed an Android application for the use of acoustic data collection. It gives the user two collection methods, manual mode or automatic collection. Both require no extra hardware for the user to start collecting the acoustic noise level around a city. The exact power being used by the application is unknown. To verify the model, a Sauter SU-130 sound level meter was used and compared with the data collected via the app. A visual dashboard is available to the end-user which displays the collected data on to a map. Another Android application was developed in~\cite{aspuru}. This application allows the user to record and visualize the acoustic comfort in an area as well as push the results to a server for analysis from the authors. The accuracy of their data was analyzed and verified in a separate study~\cite{aspuru2}.
    
    The main focus of the papers reviewed so far have been focused on developing a low power solution for ambient sound collection. The other concern this paper attempts to address is finding an effective way to collect and handle the sound data while being able to integrate it into the design. The authors in~\cite{kirillov} discuss a mobile system dedicated to the sensing of noise pollution in urban environments. It focuses on determining the level of attenuation as part of the sound collection process in addition to gathering the actual audio levels. The work in~\cite{patil} specializes in the tracking of audio data with an emphasis on vehicles. The data being sent is based on a certain threshold value which will omit any data below it. Unfortunately, no schematic of the system was made and the power consumption was not presented. This severely limits the usefulness of the claimed results and methodologies.
        
    \begin{table*}[t]  
        \renewcommand{\arraystretch}{1.9}
        \centering
        \normalsize
        \begin{tabular}{|c|c|c|c|} \hline
            \textbf{Manufacturer} & \textbf{Microphone} & \textbf{Voltage Range (V)} & \textbf{Configuration} \\ \hline
                       \multirow{2}{*}{InvenSense} & \includegraphics[width=0.65cm]{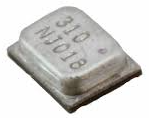} & \multirow{2}{*}{0.9 - 1.3} & \multirow{2}{*}{Analog} \\ 
             & ICS-40310~\cite{inv} & & \\ \hline
             \multirow{2}{*}{Projections Unlimited, Inc. (PUI)} & \includegraphics[width=0.85cm]{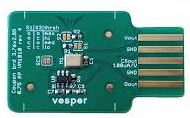} &  \multirow{2}{*}{1.6 - 3.6} &  \multirow{2}{*}{Analog} \\ 
                         & PMM-3738-VM1010-EB-R~\cite{pui} & &  \\ \hline

            \multirow{2}{*}{STMicroelectronics (ST)} &   \includegraphics[width=0.9cm]{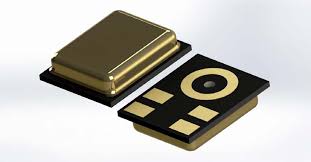} & \multirow{2}{*}{1.52 - 3.6}& \multirow{2}{*}{Analog}\\ 
                         & MP23ABS1~\cite{stanalog} &  &  \\ \hline

             \multirow{2}{*}{STMicroelectronics (ST)} &   \includegraphics[width=0.6cm]{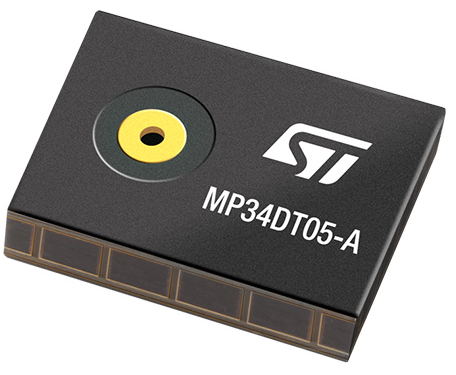}  &  \multirow{2}{*}{1.6 - 3.6} &  \multirow{2}{*}{Digital} \\ 
                        & MP34DT05-A~\cite{stdigital} &  &  \\ \hline

        \end{tabular}%
        \captionof{table}{MEMS microphone overview.}%
        \label{micinfo}%
    \end{table*}

 In~\cite{zhou}, a factory was examined and profiled based on the acoustic levels. Their goal is to reduce the amount of noise pollution by redesigning the air compressor. The work presented in~\cite{dov} highlights an improved way to collect audio data for voice activity detection. It uses both audio and visual signals with a supervised learning algorithm to detect what audio frames correspond to human voice. The authors in~\cite{zubari} use a Gaussian mixture model-based classification approach to speech detection. It's trained using spectral flow direction, a novel feature for this type of classification.  In~\cite{feki}, a multistage system is developed to analyze environmental sound and classify it. The authors claim to achieve an accuracy higher than 90\% for audio concept identification.

In comparison with these works, this paper introduces a novel method that is portable and can be implemented into many in-place solutions. Specifically, this testbed utilizes a wake-up circuit to conserve power when minimal audio activity is present, while three different technologies are examined to find the most energy-efficient.

\section{System Architecture}\label{experiments}
    The introduced system has four main components: the microphone, the circuit, the microcontroller, and the wireless unit. An overview of the system can be seen in Fig.~\ref{testbed}. A microphone collects raw audio data and then the signal is amplified. This amplified signal is then fed to both a microcontroller and a custom threshold circuit. The microcontroller only collects data when the audio threshold has been broken, then finally sends the processed data to a server via a wireless technology.
    
    \subsection{Microphone}
        Multiple microphones were considered to determine a good candidate which had a high ratio of accuracy to power. They all have varying manufacturers, voltage ranges, and configurations. A summary showing the details of each microphone can be found in Table \ref{micinfo}. While most are analog, one was picked to be digital and used as a control to compare with. The integrated ADC will lead to a higher accuracy than others, the disadvantage being a significant increase in power.
        
        The chosen microphone needed to be analog to properly interface with the threshold circuit. To make the system more sensitive to audio, an amplifier was used.
        
    \subsection{Circuit}
        The circuit that was designed has three main components: the amplifier, the envelope filter, and the threshold circuit.
    
        \subsubsection{Top level design}
\begin{figure}[t!]
                \centering    \includegraphics[width=1\columnwidth]{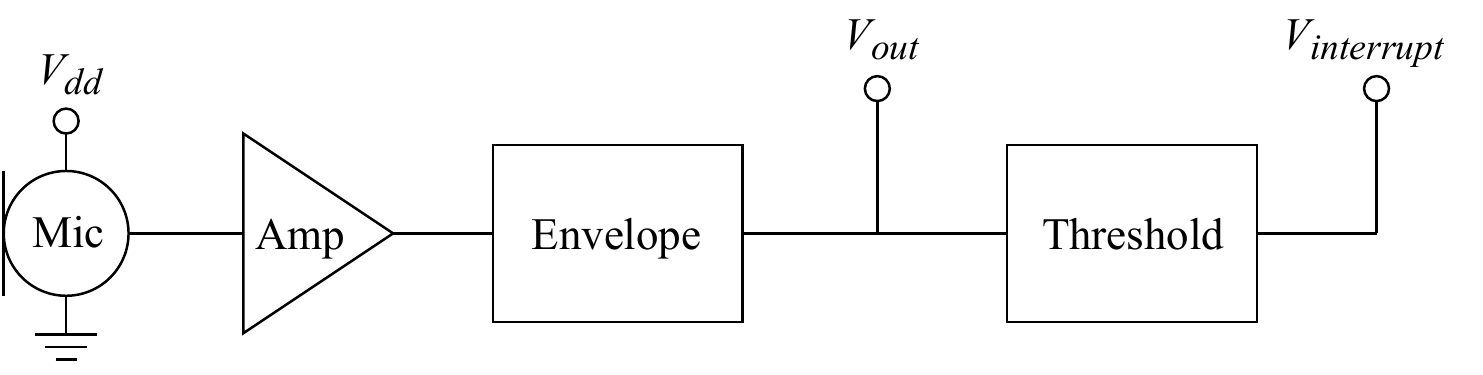}
                \caption{Top level view of analog circuit.}
                \label{top_circuit}
            \end{figure}

            Multiple circuits were designed to collect and analyze the data. An overview of the electrical components can be seen in Fig.~\ref{top_circuit}. This shows the three simplified components. An external amplifier was needed to allow the ADC to accurately capture small changes in signal amplitude, essentially increasing the resolution. The envelope filter averaged the positive amplitude using resistors and a capacitor. This is used to bypass the need for pre-processing before data transmission. The main feature of this prototype, the wake-up circuit, was designed to output a high or low value based on a threshold.
            
            %An addition had to be made when considering a system without a microcontroller, which can be seen in Fig.~\ref{top_circuit_xbee}. This shows an intermediate envelope circuit which is used to bypass the need for pre-processing before data transmission.

        \subsubsection{Amplifier}
\begin{figure}[t!]
                \centering
                \includegraphics[width=1\columnwidth]{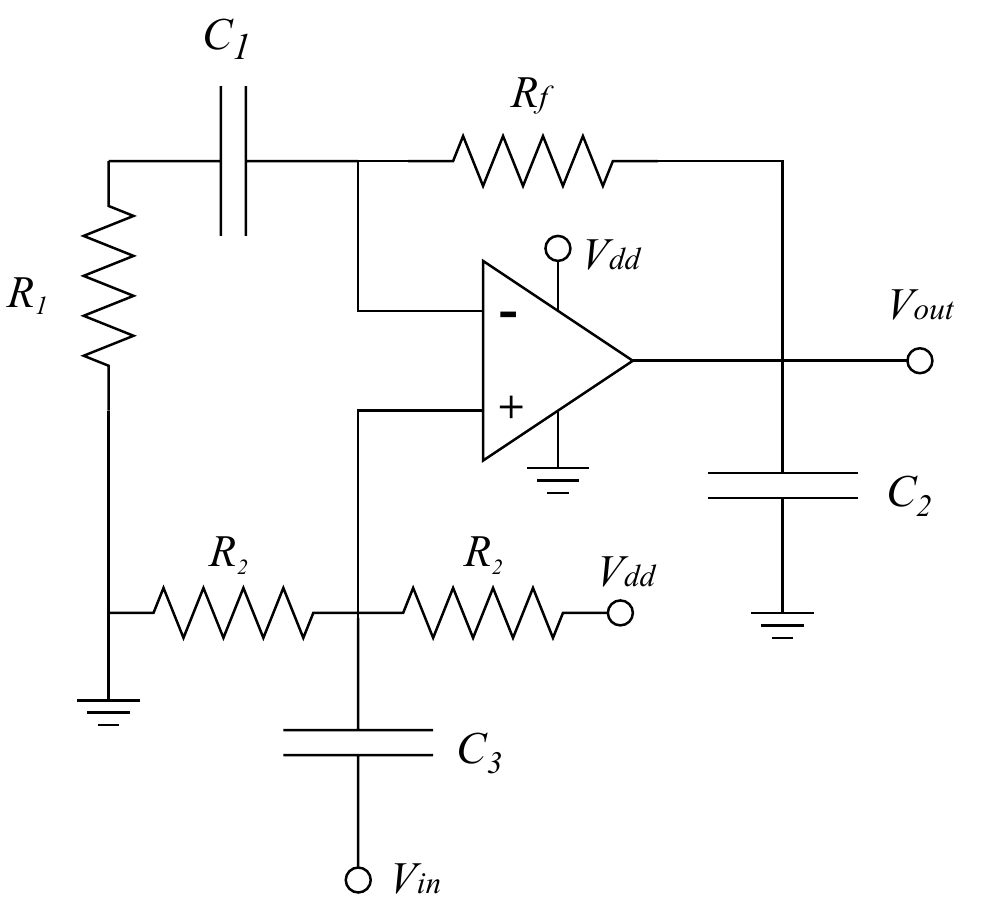}
                \caption{Amplifier circuit.}
                \label{amp_circuit}
            \end{figure}

            An amplifier was needed to make the signal recognizable to an ADC with a standard resolution. This was also important for the wake-up circuit, making it much more sensitive to changes in noise. The amplifier was designed to have low noise and a high gain. The amplifier is shown in Fig.~\ref{amp_circuit} The negative feedback adds stability to make a much cleaner signal. The gain of the amplifier can be determined from $R_f$ and $R_1$ as seen in the following:
                    
            \begin{equation}
                A_v\ =\ 1\ +\ \frac{R_f}{R_1}
                \label{eq_gain}
            \end{equation}
            
            where $A_v$ is the gain of the amplifier. This is the transfer function for a typical non-inverting amplifier design. The common-mode input and output can be seen by the following equation:
            
            \begin{equation}
                CM\ =\ \frac{V_{dd}}{2}
                \label{eq_cm_gain}
            \end{equation}
            
            where the common-mode voltage is simply the supply voltage divided by two. This is from the voltage divider defined by $R_2$.

            % \begin{figure}[t!]
            %     \centering    \includegraphics[width=1\columnwidth]{figures/top_circuit.pdf}
            %     \caption{Top level view of circuit.}
            %     \label{top_circuit}
            % \end{figure}
            
        \subsubsection{Envelope}
\begin{figure}[t!]
                \centering
                \includegraphics[width=1\columnwidth]{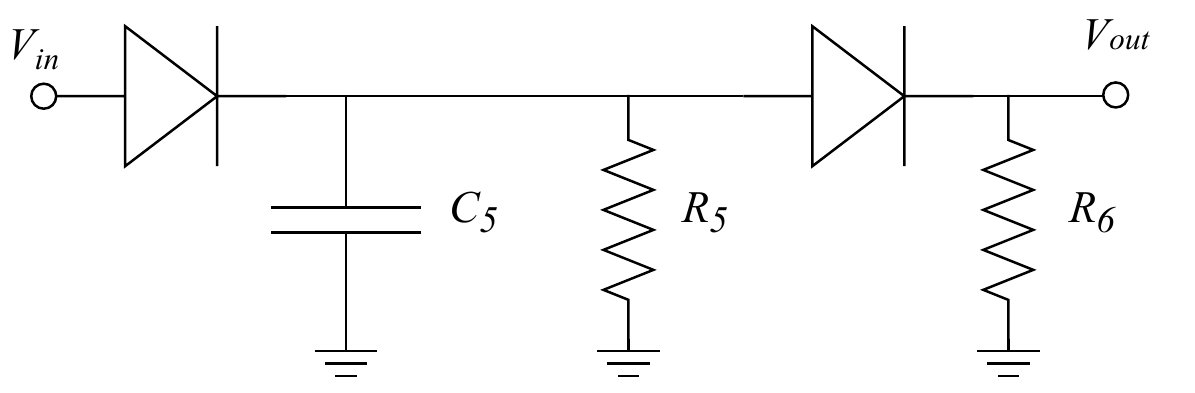}
                \caption{Envelope circuit.}
                \label{envelope_circuit}
            \end{figure}

            An envelope circuit was designed to average the magnitude of audio. The schematic of this circuit can be seen in Fig.~\ref{envelope_circuit}. This surpassed the need for pre-processing. The purpose of $C_5$ and $R_5$ is to increase the time constant, filtering the output. The second diode and $R_6$ are used to lower the common-mode voltage since the XBee ADC uses a voltage reference of 1.2 V. 
        
        \subsubsection{Threshold circuit}
\begin{figure}[t!]
                \centering                \includegraphics[width=1\columnwidth]{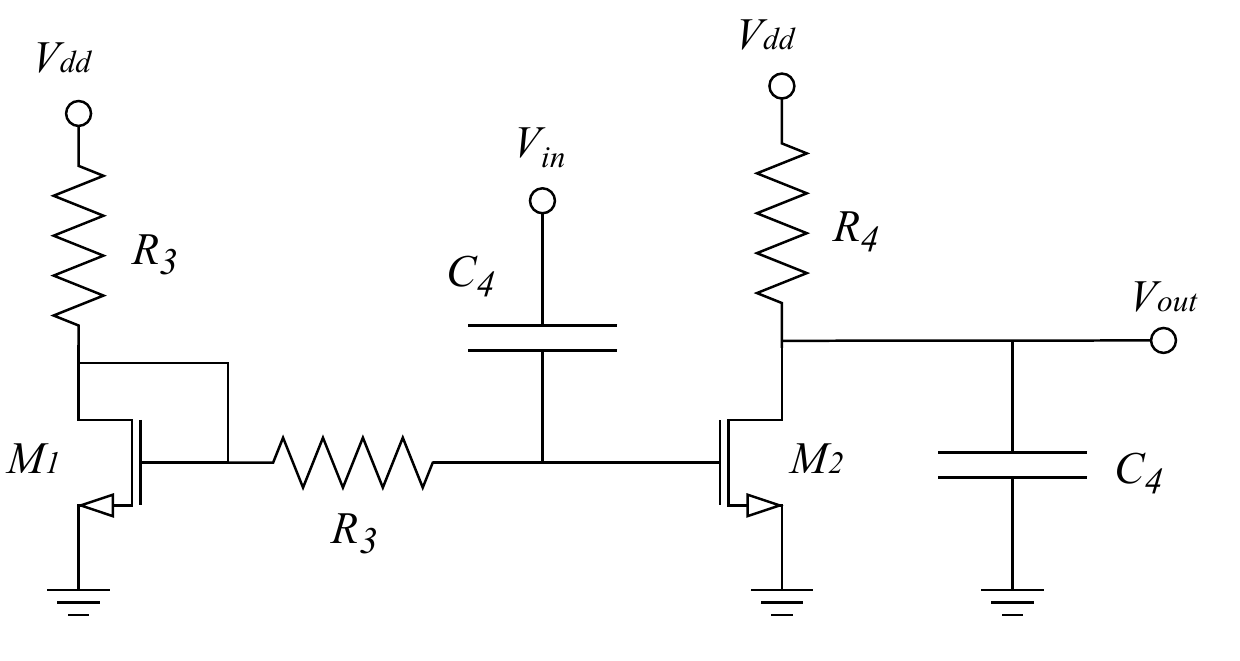}
                \caption{Threshold circuit.}
                \label{wake_circuit}
            \end{figure}

           The threshold circuit was used to control an interrupt on the microcontroller. This disabled sleep mode and allowed the microcontroller to send audio data through the various wireless technologies for a brief period before going back into sleep mode. The schematic for the wake-up circuit can be seen in Fig.~\ref{wake_circuit}.
                    
            \begin{table}[t!]
            \renewcommand{\arraystretch}{1.4}
            \centering
            \normalsize
            \begin{tabular}{|C{2.7cm}|C{2.7cm}|} \hline
                Parameter & Value\\ \hline
                $V_{dd}$ & 3.3 V\\ \hline
                $R_f$ & 100 k$\Omega$\\ \hline
                $R_1$ & 1 k$\Omega$\\ \hline
                $R_2$ & 1 M$\Omega$\\ \hline
                $R_3$ & 10 M$\Omega$\\ \hline
                $R_4$ & 500 k$\Omega$\\ \hline
                $R_5$ & 10 M$\Omega$\\ \hline
                $R_6$ & 100 k$\Omega$\\ \hline
                $C_1$ & 22 $\mu$F\\ \hline
                $C_2$ & 100 nF\\ \hline
                $C_3$ & 4.5 $\mu$F\\ \hline
                $C_4$ & 1.5 nF\\ \hline
                $C_5$ & 9 $\mu$F\\ \hline
                \end{tabular}%
                \captionof{table}{Circuit parameters.}%
                \label{parameters}%
            \end{table}
        
            Two transistors are used, $M_1$ regulates the DC offset of the input, $M_2$ is biased in such a way that it will saturate when the input is high. Once $M_2$ is saturated, it ties the output to ground. Alternatively, $V_{out}$ will be close to $V_{dd}$ when $M_2$ is in the cut-off region. This means that the output will be grounded when the threshold is broken and $V_{dd}$ when no significant noise is heard. At this time, the threshold can only be configured through hardware but may be configurable in future iterations.
                    
            All of the circuit parameters used for this set of experiments can be found in Table \ref{parameters}. Using Eq. (\ref{eq_gain}), the gain can be calculated based on these parameters:
            
            \begin{dmath}
                A_v\ =\ 1\ +\ \frac{100 k \Omega}{1 k \Omega}\ \hiderel{=}\ 101\ \frac{V}{V}
            \end{dmath}
               
            The gain becomes 101 $\frac{V}{V}$, amplifying the signal by approximately 20 dB. This is necessary for the ADC to properly recognize relatively small changes in the ambient acoustic noise.

    \subsection{Microcontroller}
        A microcontroller was needed to correctly process the analog data incoming from the amplifier. The NodeMCU was picked for its low power, flexibility, and ability to harness many types of wireless technologies. The NodeMCU uses an ESP8266 microprocessor with a 10-bit ADC.
        
        %Once the data had been converted by the ADC, the microcontroller performed an algorithm to convert raw audio to a magnitude that can be found in Algorithm~\ref{algo_level}. This involved sampling and tracking the maximum value of all samples.
        
        % \begin{algorithm}[t!]
        %     \normalsize
        %     \caption{: Decide the noise level}
        %     \begin{algorithmic}
            
        %     \State{averageSize = 30}
        %     \State{sampleSize = 50}\\

        %     \For {i $<$ averageSize}
        %         \State{audioMax = 0}\\
                
        %         \For {j $<$ sampleSize}
        %             \State{val = analogRead(0)}\\
                    
        %             \If {audioMax == val}
        %                 \State{audioMax = val}
        %             \EndIf
        %         \EndFor
        %         \State{averageMag += audioMax}
        %     \EndFor
            
        %     \State{averageMag /= averageSize}
        
        %     \end{algorithmic}
        %     \label{algo_level}
        % \end{algorithm}
        
        %The microcontroller averaged the max values recorded to form the correct trend. Essentially, this makes a higher value when the environment is louder, and a lower value with the absence of sound.
        
        When the microcontroller was not in sleep mode, it connected to the server using either Wi-Fi (2.4 GHz), BLE 4.0, or Zigbee. The microcontroller then slept if the noise threshold was not broken. The XBee S2C is capable of converting an analog value to digital and transmitting the data using Zigbee. This bypasses the need for a microcontroller and will be tested alongside the microcontroller variant of the prototype.
    
    \subsection{Wireless technology}
        Three wireless technologies were tested to determine which was the lowest power, and show the sleep functionality in a variety of cases.
        
        %\subsubsection{Wi-Fi (2.4 GHz)}
        \subsubsection{Wi-Fi}
            Wi-Fi is a common technology found in most homes. It allows users to connect to the internet wirelessly by using an access point. The power consumption and bandwidth are typically high, keeping up with the needs of the consumer \cite{fallis}. For the application of an audio acquisition system, the bandwidth should not be as much of an issue when compared to the power being expended.
        
            The NodeMCU has an integrated proprietary Wi-Fi (2.4 GHz) chip, giving it the capability to transmit data. It was able to directly send data to a server.
        
        %\subsubsection{Bluetooth Low Energy (BLE) 4.0}
        \subsubsection{Bluetooth}
            Bluetooth is a popular technology used for many wireless devices. It is designed to be a low-power solution that conserves energy to promote battery life in electronics. The power is typically less than Wi-Fi, but the throughput suffers~\cite{fallis}.
            
            The DSD TECH HM-10 was used to connect the microcontroller using BLE 4.0~\cite{hm10}. It communicates with a separate device to wirelessly transmit the data.
        
        \subsubsection{Zigbee}
            Zigbee is a lesser-known protocol capable of low power transmission, meant to be used specifically for IoT (Internet of Things)~\cite{zigbee}. This makes it a good match for the application of low power collection of audio. 
            
            The XBee S2C module was being used \cite{xbee}. This provided a platform for the microcontroller to transmit using the Zigbee protocol. Similar to Bluetooth, the Zigbee experiments required a secondary unit with a matching XBee to provide wireless communication.
            
            An advantage of the XBee is the integrated ADC, which allows for wireless transmission without the need for an external microcontroller.

\section{Experimental Procedure}\label{procedure}

    \subsection{Microphone evaluation}
        To examine which microphone was the most accurate, the coherence between the original waveform and the recorded waveform was found. This shows how similar the recording is to the original signal in terms of the frequency response. Magnitude squared coherence was used on the pre-processed signals to create an output which shows how much the frequency domains of the signals match.
        
        The coherence was conducted in Matlab by using functions in the signal processing library. A procedure was used to filter the two signals and end up with a percentage value that represents how closely the recording resembles the source. The simple list of steps includes:
            
        \begin{itemize}
        \setlength\itemsep{0.5em}
            \item Down-sample all recordings to 8 kHz.
            \item Lineup source and recording.
            \item Limit time to exactly 80 seconds.
            \item Find magnitude squared coherence.
            \item Take envelope of the signal.
            \item Average the envelope.
        \end{itemize}
            
        The purpose of down-sampling is to simplify the data and give every dataset a fair comparison without removing the frequencies that matter as well as increase the performance of the script. This uses decimation to make the signal simpler, and thus easier to compare.
        
        Although the magnitude squared coherence will convert the signals to the frequency domain, it is still important to properly line up the source and recording. This eliminates any frequencies that happened before or after the recording interval. The offset can be found by comparing the similarities of the two signals and performing a correlation that is much more basic when compared to the magnitude squared coherence. This offset was used to perfectly line up the signals by trimming the microphone clip to exactly 80 seconds. The offset was based on a 10 second clip of the source and recording which was shown to be enough to properly line up the signals. The magnitude squared coherence finds the cross power spectral density in the numerator which can be seen below:
            
        \begin{equation}
            C_{xy}(f) = \frac{|G_{xy}(f)|^2}{G_{xx}(f)G_{yy}(f)}
        \end{equation}
       where $G_{xy}$ is the cross-spectral density defined by:
    
        \begin{equation}
            G_{xy}(f) = \int_{- \infty}^{\infty} R_{xy}(t)e^{-j \omega t} dt
        \end{equation}
        where $R_{xy}$ is the cross-correlation of the two original signals, $x(t)$ and $y(t)$. The cross-correlation of the two signals can be described as:
        
        \begin{equation}
            R_{xy}(t) = [\overline{x(-t)}*y(t)](t)
        \end{equation}
    which involves the two signals being convoluted. 
    
    The envelope of this attempts to create a coherence that more accurately represents how closely the two signals match, this finds the outline of the signal. Finally, the mean is taken of the resulting envelope to determine the accuracy value. The code showing all the modifications to the signal can be seen in Algorithm \ref{algorithm}.
 
        \begin{algorithm}[t!]
            \caption{: Find the effective coherence value}
            \begin{algorithmic}

            \State{y\_source = resample(y\_source, 8000, Fs\_source)}
            \State{y\_mic = resample(y\_mic, 8000, Fs\_mic)}\\
            
            \State{y\_source\_clip = y\_source(1:8000*10)}
            \State{y\_mic\_clip = y\_mic(1:8000*10)}\\
            
            \State{mic\_delay = finddelay(y\_source\_clip, y\_source\_clip)}
            \State{y\_mic = y\_mic(mic\_delay:mic\_delay + 8000*80 - 1)}\\
            
            \State{[corr, f] = mscohere(y\_source, y\_mic, [], [], [], 8000)}\\
            
            \State{corr = envelope(corr, 100, 'peak')}\\
            
            \State{corr\_value = mean(corr)}
        
            \end{algorithmic}
            \label{algorithm}
        \end{algorithm}
        
        %The Urban Sounds data set was used, this contains 4 second clips from categories including human voice, music, mechanical, and nature sounds \cite{urbansound}. 
        
        The experimental conditions were  kept constant through all the experiments. The list of conditions for the experiment include:

        \begin{itemize}
        \setlength\itemsep{0.5em}
            \item 80 second audio clip of various urban sounds.
            \item Nexus 5 at 53.3\% volume, connected to speakers.
            \item Audio clips out of speakers are 96 kHz.
            \item Distance of 30 cm from speaker and microphone.
        \end{itemize}

        A Nexus 5 was used to play the 80 seconds clip for each experiment. This amount of time was chosen to get an average accuracy for each microphone, reducing the effect of any outliers. The Nexus 5 was connected to a pair of Nex Tech analog speakers to provide a much higher quality sound when compared to the embedded speakers on the Nexus 5 \cite{speakers}. This was kept at a constant 30 cm away from the respective microphone which ensured consistent data. Each microphone was pointed directly at the speakers for a fair comparison. The volume of the Nexus 5 was set to a constant of $\frac{8}{15}$ bars, or 53.3\% while the speakers were kept at a maximum gain.

        The Urban Sound dataset was used to test with, it contains audio categorized into four main sections~\cite{urbansound}. This includes human, nature, music, and mechanical based sounds. Twenty seconds of each category was used for testing, totaling 80 seconds for each data point. The dataset features multiple sampling rates, channels, and resolution. To normalize this data, all clips were up-sampled to 96 kHz and each clip was limited to one channel, also known as mono. This was done with the Sox Linux tool by utilizing linear interpolation to reduce any inaccuracies during the coherence process. Down-sampling has the potential to remove data points so it was excluded as an option.
        
    \subsection{Power evaluation}
        It is important to measure the power savings created by the wake-up functionality. To do this, the Monsoon power monitor was being used. It is capable of providing very accurate measurements of energy consumption~\cite{monsoon}.
        
        Measuring the power consumption of both the sleep and active modes of each permutation of the prototype will show 
        
        The Urban Sounds data set was being used for testing. Audio clips were created and designed to simulate a realistic outdoor environment. Specifically, 20 seconds of audio from a category was played, followed by 100 seconds of silence. This was repeated for each category, totaling 8 minutes of audio. The clips were organized by loudness to test the power consumption at different noise levels.

    \subsection{Calibration}
        The voltage being recorded by the microcontroller ADC does not mean much until it can be related to a more accurate measuring device. Putting data into a human-readable format is important both for interpretation and comparison to other data sets.
    
        A handheld Sound Pressure Level (SPL) meter was used to calibrate the prototype. An equation was formed to convert values on the ADC to one representing decibel noise levels. The Decibel X Android app was also calibrated and used to measure the average noise level when recording and testing with the prototype. The application was seen to only have an error of approximately $\pm$2 dB.

\section{Results}\label{results}

    \subsection{Microphone selection}
        After the magnitude-squared coherence was conducted, the accuracy values could be analyzed as well as the power dissipation. As shown in Table~\ref{accuracy}, it is clear that the InvenSense has the lowest power consumption. This is partially because the supply voltage is 1 V as opposed to all the other microphones which have a nominal voltage of 3.3 V. However, the disadvantage is requiring a voltage regulator to step down from 3.3 V. This would require an excessive amount of power.

        \begin{table}[t!]
            \renewcommand{\arraystretch}{1.7 }
            \centering
            \normalsize
            \begin{tabular}{|C{2cm}|C{2cm}|C{2cm}|} \hline
                \textbf{Microphone} & \textbf{Power (mW)} & \textbf{Accuracy}\\ \hline
                InvenSense & 0.0079 & 0.7359\\ \hline
                PUI & 0.3545 & 0.6966\\ \hline
                ST (Analog) & 0.3480 & 0.7187\\ \hline
                ST (Digital) & 2.0813 & 0.7577\\ \hline
            \end{tabular}%
            \captionof{table}{Microphone accuracy and power.}%
            \label{accuracy}%
        \end{table}
                            
        \begin{figure}[t!]
            \centering
            \includegraphics[width=1\columnwidth, trim={100 230 100 255}, clip]{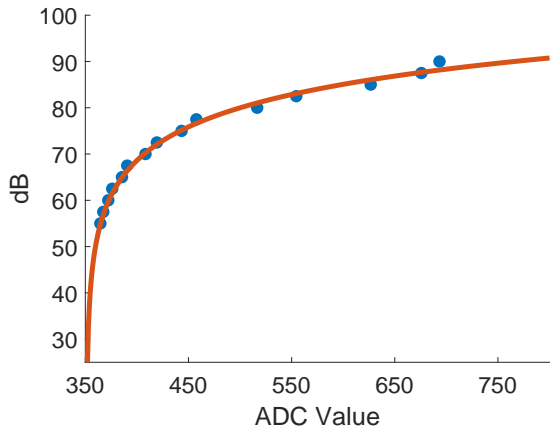}
            \caption{SPL calibration curve.}
            \label{calibration}
        \end{figure}
        
        \begin{equation}
            dB = -290.5(x-350)^{-0.04258} + 314.7
        \end{equation}
        
        \begin{table*}[t]
            \renewcommand{\arraystretch}{1.4}
            \centering
            \normalsize
            \begin{tabular}{|C{3.5cm}|C{2.5cm}|C{2.5cm}|} \hline
                \textbf{Component} & \textbf{Power (mW) [Transmit]} & \textbf{Power (mW) [Sleep]}\\ \hline
                Microphone & 0.35 & 0.35\\ \hline
                Amplifier & 0.20 & 0.20\\ \hline
                Threshold & 0.07 & 0.07\\ \hline
                %Arduino & 268.20 & 208.54\\ \hline
                NodeMCU & 91.84 & 16.76\\ \hline
                %CC3000 Wi-Fi (2.4 GHz) & 531.34 & 32.26\\ \hline
                Wi-Fi (2.4 GHz) & 265.75 & 0.00\\ \hline
                BLE 4.0 & 49.02 & 8.47\\ \hline
                Zigbee & 33.68 & 0.07 \\ \hline
            \end{tabular}%
            \captionof{table}{Power results by component.}%
            \label{power_static}%
        \end{table*}

        The highest accuracy is the digital ST microphone. This makes sense as the integrated ADC is configured to work specifically for the microphone. Consequently, it also has the highest power consumption. Additionally, exclusively using a digital microphone would prohibit the use of the threshold circuit since the output of the microphone is PDM rather than a pure analog signal. The analog ST microphone shows high accuracy and low power, working without the need for a voltage regulator. These reasons are why the analog ST microphone was chosen as the best candidate for the remaining experiments.

    \subsection{Decibel calibration}        
        A handheld SPL meter was used to calibrate the data being transmitted from the microcontroller. An equation was formed to transfer the voltages being reported to the server into a Z-weighted decibel value. The visual calibration can be found in Fig.~\ref{calibration}. The $R^2$ value is 0.995, indicating a very strong correlation. The equation is as follows:

        \begin{figure*}[t!]
            \centering
            \subfloat[\normalsize Sleep]{\includegraphics[width=0.85\columnwidth, trim={95 245 95 245}, clip]{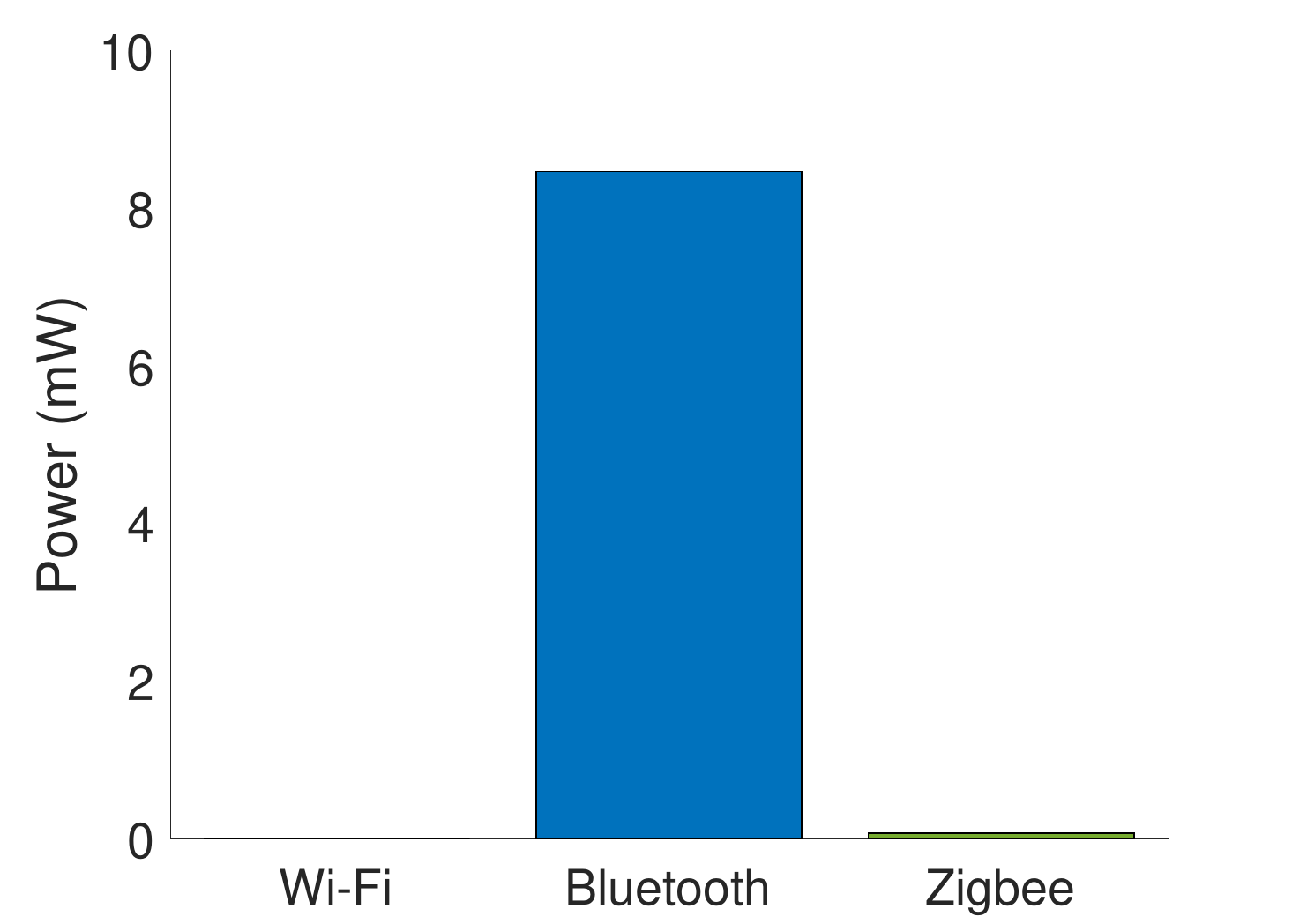}\label{wireless_power_sleep}}
            \subfloat[\normalsize Transmit]{\includegraphics[width=0.85\columnwidth, trim={95 245 95 245}, clip]{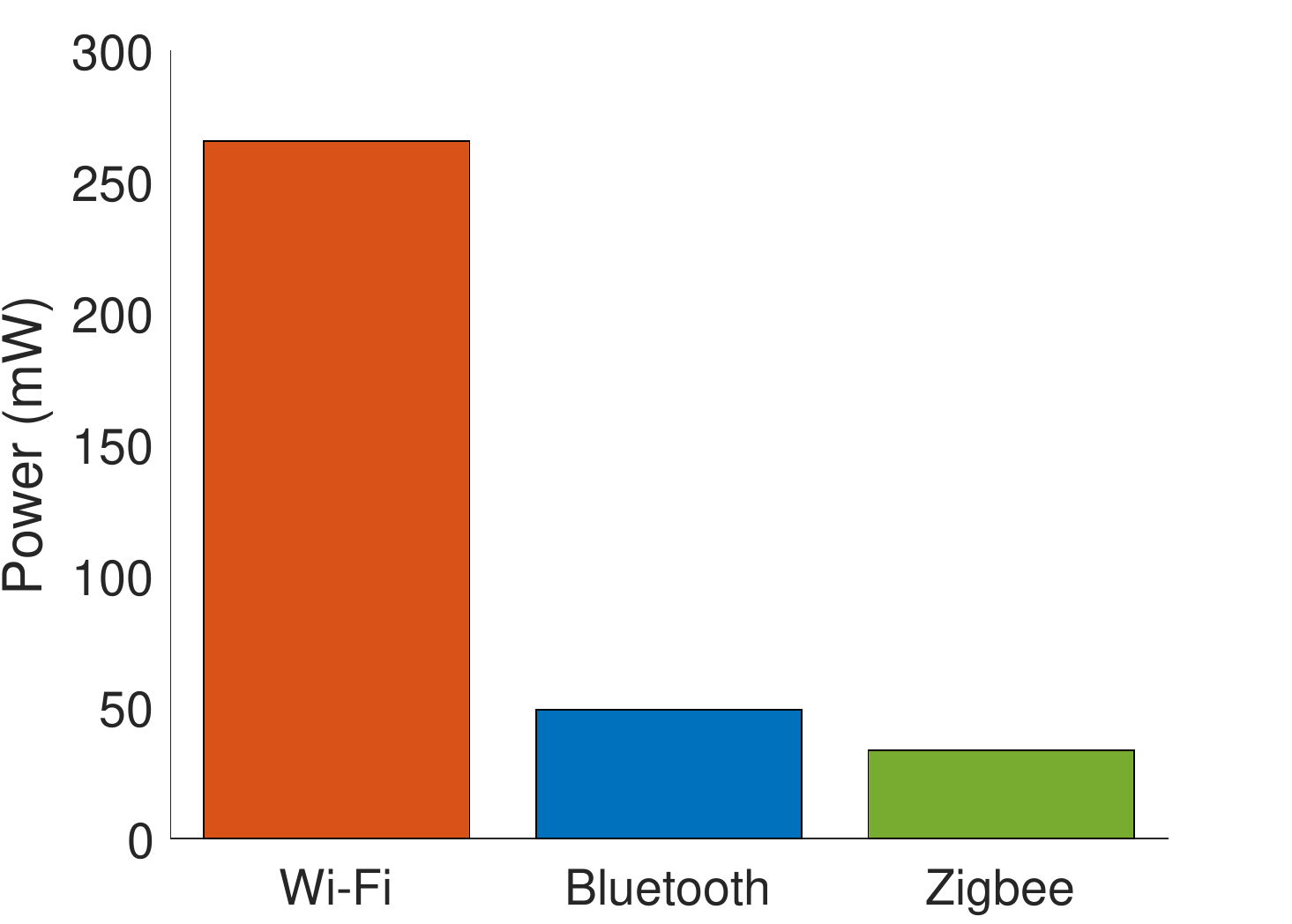}\label{wireless_power}}
            \caption{Wireless power consumption.}
        \end{figure*}

        \begin{table*}[t!]
            \renewcommand{\arraystretch}{1.4}
            \centering
            \normalsize
            \begin{tabular}{|C{3cm}|C{3cm}|C{3cm}|C{2cm}|} \hline
                \textbf{Wireless Technology} & \textbf{Power (mW) [Sleep]} & \textbf{Power (mW) [Transmit]} & \textbf{Savings ($\%$)}\\ \hline
                Wi-Fi (2.4 GHz) &  16.76 & 357.59 & 95.3\\ \hline
                BLE 4.0 & 25.23 & 140.86 & 82.1\\ \hline
                Zigbee & 16.83 & 160.43 & 89.5\\ \hline
                Zigbee (Standalone) & 1.00 & 34.30 & 97.1\\ \hline
            \end{tabular}%
            \captionof{table}{Total power results of prototypes.}%
            \label{power_wireless}%
        \end{table*}

        where $x$ is the ADC value after processing on the microcontroller. This power trend makes sense as decibels are logarithmic, meaning it should follow an exponential trend.

    \subsection{Power consumption}
        The power consumption of the prototype was measured with the Monsoon power monitor. Each module was considered individually to analyze the biggest power draw in the system. The power consumption of each component can be seen in Table~\ref{power_static}. As expected, the microcontroller and Wi-Fi module have the largest power draw by a very significant amount when transmitting. Both Bluetooth and Zigbee are comparable in terms of power, though the power of the XBee is slightly lower, especially when sleeping. The ESP8266 Wi-Fi module of the NodeMCU was considered to have a power of zero when sleeping since it's integrated into the microcontroller itself.

        % \begin{table}[t]
        %     \renewcommand{\arraystretch}{1.7}
        %     \centering
        %     \normalsize
        %     \begin{tabular}{|C{0.8cm}|C{1.9cm}|C{1.9cm}|C{1.9cm}|} \hline
        %         \textbf{dB} & \textbf{Power (mW) [Wi-Fi (2.4 GHz)]} & \textbf{Power (mW) [BLE 4.0]} & \textbf{Power (mW) [Zigbee]}\\ \hline
        %         65.1 & 325.5 & 232.71 & 228.01\\ \hline
        %         71.6 & 331.3 & 233.75 & 229.25\\ \hline
        %         77.6 & 336.6 & 234.70 & 230.37\\ \hline
        %     \end{tabular}%
        %     \captionof{table}{Trial power results.}%
        %     \label{power_trials}%
        % \end{table}

        The power of each wireless technology was specifically compared in Fig. \ref{wireless_power_sleep} and Fig. \ref{wireless_power}. These figures show a comparison of each wireless technology in both the sleep mode and transmitting mode. A clear trend of Wi-Fi consuming significantly more power than the other two technologies can be seen. Zigbee uses the least while BLE is in the middle.

        Table \ref{power_wireless} puts each technology in perspective in terms of the total power being saved by the sleep functionality of the system. The most apparent savings occur when using Wi-Fi since the active power of the Wi-Fi module is so much greater than the sleeping power. The standalone system consisting of the XBee without a microcontroller has the lowest overall power. The increase in sleeping power is most likely due to extra leakage current from the use of the ADC on the XBee. 

        % Three trials using 480 seconds of audio each were performed, each with a different average dB value. The summary of power measurements can be found in Table~\ref{power_trials}. Initially, Wi-Fi was tested, the other technologies were derived by swapping in the power of each technology and calculating the power based on the transmit/sleep duty cycle. Based on this logic, it is possible to make an equation that can be generalized to any system adopting an analog threshold circuit design. This generalized equation can be seen below:
        
        % \begin{equation}
        %     P = (fraction_{on})(P_{on}) + (fraction_{sleep})(P_{sleep})
        %     \label{eq_power}
        % \end{equation}
        
        % Considering that the prototype is currently configured to sleep 3 seconds after the threshold is passed, the total time on is 23 seconds per category. Each category totals to 120 seconds, so the amount of silence is 97 seconds per category. This can be used to find the theoretical power of the Wi-Fi prototype in Eq. (\ref{eq_power}) like so:

        % \begin{dmath}
        %     P = \frac{23\ s}{120\ s}*800.16\ mW + \frac{97\ s}{120\ s}*241.42\ mW\\\\
        %     =\ 348.51\ mW
        % \end{dmath}
        
        %Looking back at Table \ref{power_trials}, the maximum power for Wi-Fi is only 336.6 mW. When compared to the estimate, this dispute can be explained by the fact that not every audio clip triggers the threshold of the prototype. Regardless, the estimate is very close to the actual value. 
        
    \subsection{Comparison with similar approach}
        To prove the relevance of the power savings achieved, an in-place solution will be compared and analyzed. This will be compared to the Zigbee standalone implementation in this paper since it has the lowest overall power consumption. When looking at the work in~\cite{risojevic}, the average power consumption claimed is 45 mW. This is based on a periodic interrupt to transition from sleep to active, a common technique used to save power in such systems. If a quiet environment is considered, no significant sound would be reported. In this situation, the average would still be 45 mW in the work being compared while the presented system will use 1 mW. This leads to a power savings of over 97\%. The paper being compared assumes the prototype has a battery size of 2900 mAh which means their prototype can last for 9 days based on the average power consumption. The amount of improvement shown in this prototype leads to the system lasting approximately 400 days when no significant audio is detected. 
        
        If a moderately noisy environment is considered, this will change the average power of the presented system. The system presented system will never use more power than the system being compared since a 100\% duty cycle would lead to a power of only 34 mW. This means that in the worst case, the prototype will last just over 11 days. Based on this result, the improved version presented in this paper will always outperform that of the comparison work.
        
        Another big advantage of the presented system is that it won't miss significant events, unlike a periodic sampling system. This is because the threshold circuit will always activate the transmission of data once the noise threshold has been reached. The envelope circuit will also mean the sound that triggers the threshold will also be accounted for when reading the sample. Improving the active standard power consumption of the prototype by removing the microcontroller led to a significantly improved power consumption, meaning that in any state it will be more efficient than the comparison system. The implementation of the threshold circuit further improved the power by allowing the prototype to sleep in most cases, thus improving leading to a much more efficient prototype overall.

\section{Conclusion}\label{conclusion}
    In this work, a wireless ambient sound collection system was designed and tested. The novelty is a threshold circuit that only wakes the digital components upon a threshold being broken. This lowers the average power consumption by 97$\%$ when using the lowest power permutation of the prototype is used for data transmission, making a full-scale wireless microphone array much more feasible in smart cities. When compared to a system of similar performance, the presented system lowers power consumption. In comparison to a periodic sleep cycle, this solution ensures that relevant audio data is collected. This can also be tailored to regulate an environment that is meant to stay below a certain level of dB.
             
    The work presented is a flexible solution that can easily be tailored to other audio collection systems. Three wireless technologies were used to both measure the power consumption and find the effectiveness of a threshold circuit in each. In the future, large-scale wireless audio collection systems will be possible, this is one step towards a power-efficient system capable of operating for months with no human maintenance. 
    
    %There are various ways to improve the prototype in the future. Selecting a more power-efficient microcontroller would improve battery life substantially. 
    
    Testing a greater variety of MEMS microphones could also potentially improve accuracy and power. Developing an analog A-weighted filter would make the noise level better represent the human ear when collecting data. Developing a dynamic threshold circuit to regulate the duty cycle would help reduce power further.

\section*{References}
\bibliography{ElsIoT}

\end{document}